%% file: main.tex
\documentclass[aps,prl,twocolumn,
			   groupedaddress,superscriptaddress,
			   amsfonts,amssymb,amsmath,
			   citeautoscript,longbibliography,
			   a4paper
			   ]{revtex4-2}

\input{setup}

\graphicspath{{figs/}}

\setboolean{togglecomments}{true}
\setboolean{toggletodos}{true}
\setboolean{togglechanges}{false}

\begin{document}
\title{Prevalence of two-dimensional photonic topology}


\newcommand{\mitaffil}{Department of Physics, Massachusetts Institute of Technology, Cambridge, MA 02139, USA}
\newcommand{\pennaffil}{Department of Physics, The Pennsylvania State University, University Park, PA 16801}
\newcommand{\princetonaffil}{Department of Physics, Princeton University, Princeton, NJ 08542, USA}
\newcommand{\emoryaffil}{Department of Physics, Emory University, Atlanta, GA 30322, USA}
\newcommand{\dtuaffil}{Department of Photonics and Electrical Engineering, Technical University of Denmark, Lyngby 2800, Denmark}

\author{Ali Ghorashi}
\email{aligho@mit.edu}
\affiliation{\mitaffil}

\author{Sachin Vaidya}
\affiliation{\mitaffil}
\affiliation{\pennaffil}

\author{Mikael Rechtsman}
\affiliation{\pennaffil}

\author{Wladimir Benalcazar}
\affiliation{\emoryaffil}

\author{Marin Solja\v{c}i\'c}
\affiliation{\mitaffil}

\author{Thomas Christensen}
\email{thomas@dtu.dk}
\affiliation{\mitaffil}
\affiliation{\dtuaffil}

\keywords{}
\pacs{}

\begin{abstract}    
    The topological characteristics of photonic crystals have been the subject of intense research in recent years.
    Despite this, the basic question of whether photonic band topology is rare or abundant---i.e., its relative prevalence---remains unaddressed.
    Here, we determine the prevalence of stable, fragile, and higher-order photonic topology in the 11 two-dimensional crystallographic symmetry settings that admit diagnosis of one or more of these phenomena by symmetry analysis.
    Our determination is performed on the basis of a data set of \num{550000} randomly sampled, two-tone photonic crystals, spanning 11 symmetry settings and 5 dielectric contrasts, and examined in both transverse electric (TE) and magnetic (TM) polarizations.
    We report the abundance of nontrivial photonic topology in the presence of time-reversal symmetry and find that stable, fragile, and higher-order topology are generally abundant.
    Below the first band gap, which is of primary experimental interest, we find that stable topology is more prevalent in the TE polarization than the TM; is only weakly, but monotonically, dependent on dielectric contrast; and that fragile topology is near-absent.
    In the absence of time-reversal symmetry, nontrivial Chern phases are also abundant in photonic crystals with 2-, 4-, and 6-fold rotational symmetries but comparatively rare in settings with only 3-fold symmetry.
    Our results elucidate the interplay of symmetry, dielectric contrast, electromagnetic polarization, and time-reversal breaking in engendering topological photonic phases and may inform general design principles for their experimental realization.
\end{abstract}
\maketitle


%
Topological properties of photonic crystals (PhCs)~\cite{lu2014topological, ozawa2019topological} have attracted substantial interest since the prediction of topological insulators in electronic systems~\cite{fu2007topological, kane2005quantum}.
In particular, recent research has run the gamut of time-reversal ($T$) invariant and broken topology, ranging across topological degeneracies~\cite{lu2013weyl}, Chern phases~\cite{haldane2008possible, wang2009observation, skirlo2015experimental}, higher-order topology~\cite{he2020quadrupole, xie2018second}, and beyond.
Here, we investigate the prevalence of stable, fragile, and higher-order photonic topology in all two-dimensional (2D) symmetry settings that admit their symmetry-based identification.
In this pursuit, we create a data base of topological 2D PhCs, analogous to recent data bases of electronic and phononic materials~\cite{zhang2019catalogue, tang2019comprehensive, vergniory2019complete, vergniory2022all, sodequist2022abundance, xu2022catalogue}.

Our work is motivated by the following simple questions:
How prevalent is topology in 2D PhCs in a statistical sense?
How is this prevalence affected by dielectric contrast, symmetries, and mode polarization?
Are certain regions of the photonic band structure more likely to be topological than others?
Answering these questions is of fundamental importance for the understanding, design, and experimental realization of photonic topological phenomena, because an understanding of the parameter space of topology can inform general design principles---e.g., the symmetries, geometric features, and range of dielectric contrasts most advantageous to engendering topological phases. 
To address these questions, we adopt the recent frameworks of topological quantum chemistry (TQC)~\cite{cano2021band, bradlyn2017topological} and symmetry indicators~\cite{kruthoff2017topological, po2017symmetry, po2020symmetry}, which have recently seen application and adaption also to the photonic context~\cite{de2019engineering, christensen2022location, kim2023automated, vaidya2023topological, vaidya2023response, antonio2023transversality, chiara2023axion}.
Their highly efficient and broadly-applicable qualities naturally lend them to high-throughput calculations, enabling us to answer the above questions by directly and comprehensively sampling the design space of 2D PhCs.

\paragraph{Methods}
We briefly summarize the salient ideas of the symmetry-based approach, which diagnoses topology from the ability to decompose a set of bands into linear combinations of so-called elementary band representations (EBRs)~\cite{zak1982band}.
These EBRs span the space of symmetry-respecting atomic limits (symmetric, exponentially localized Wannier orbitals).
Conceptually, if a set of bands---a multiplet---cannot be expressed as a ``stack'' of EBRs, there is a topological obstruction to smoothly deforming the multiplet to an atomic limit.
More technically, given a trivial multiplet $\mathfrak{n}$ in a PhC with plane group symmetry $G$, $\kv$-wise gapped from all other bands, there exists a symmetry- and gap-preserving equivalence relation between $\mathfrak{n}$ and a decomposition in the EBRs of $G$~\cite{po2017symmetry,bradlyn2017topological,kruthoff2017topological}:
\begin{equation}
    \mathfrak{n}
    \sim
    \bigoplus_{\qv\alpha} m_{\qv}^{\alpha} (\mathbf{q}|D_\qv^{\alpha}).
    \label{eq:decompose_strict}
\end{equation}
Here, $(\qv|D_{\qv}^{\alpha})$ denotes an EBR induced from a maximal Wyckoff position $\qv$ in $G$, transforming like the $\alpha$th irreducible representation (irrep) $D_{\qv}^{\alpha}$ of the associated site-symmetry group $G_{\qv} \equiv \{g \in G \mid g\qv = \qv\}$, and $m_{\qv}^{\alpha} \in \{0,1,2,\ldots\}$ is the decomposition's EBR multiplicity.

\begin{figure}
    \centerline{%
    \includegraphics[scale=1]{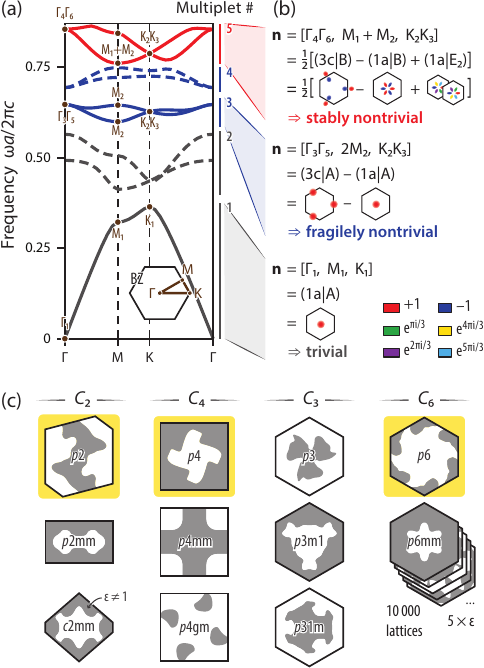}}
    \caption{%
        \textbf{Methodology, workflow, and data set}
        (a)~Band structure of a \emph{p}6-symmetric PhC, hosting 5 separable bands, i.e., multiplets (TE polarization, unit cell from (c)).
        (b)~For each multiplet, we decompose the symmetry vectors, $\nv$, in EBRs to obtain a symmetry-based diagnosis of band topology from the decomposition's coefficients.
        (c)~Examples of PhC unit cells in our data set, spanning 11 plane groups, each labelled in Hermann--Mauguin notation~\cite{ITA:6} and grouped by their rotational symmetry ($C_n$).
        Groups that admit stable, $T$-invariant topology are highlighted in yellow.
        }
    \label{fig:Workflow}
\end{figure}

The key simplification of symmetry-based frameworks~\cite{po2017symmetry,bradlyn2017topological,kruthoff2017topological} is to relax the strict functional decomposition of \cref{eq:decompose_strict} to a decomposition of band symmetries.
In detail, we decompose the so-called symmetry vector $\mathbf{n}$ of $\mathfrak{n}$---enumerating the symmetry content of $\mathfrak{n}$ through the multiplicities $n_{\kv}^{\alpha}$ of little group irreps $D_{\kv}^\alpha$ across all high-symmetry $\kv$-points---into the symmetry vectors of $(\qv|D_\qv^{\alpha})$---which, to ease notation, we shall denote by the same symbol---according to:
\begin{equation}
    \mathbf{n}
    =
    \sum_{\qv\alpha} c_{\qv}^{\alpha} (\qv|D_\qv^{\alpha}).
    \label{eq:decompose_symmetry}
\end{equation}
The symmetry-inferrable topological diagnosis is made on the basis of the decomposition coefficients $c_{\qv}^{\alpha}$~\cite{po2017symmetry, elcoro2020application, song2020fragile} (\cref{fig:Workflow}ab):
if all $c_{\qv}^{\alpha}$ can be chosen as non-negative integers, $\mathfrak{n}$ is compatible with the symmetries of an atomic limit, \ie nominally topologically trivial~%
    \footnote{While a non-negative integer decomposition in \cref{eq:decompose_strict} is a strict definition of topological triviality, the same is not true for the decompositions in \cref{eq:decompose_symmetry}.
    Rather, such a decomposition only implies compatibility with the band symmetries of a trivial atomic limit, not necessarily trivial band topology~\cite{po2017symmetry}: we call this ``nominally trivial''.
    E.g., in $T$-broken PhCs with $C_n$ symmetry, a symmetry-based diagnosis of triviality implies only that the Chern number vanishes modulo $n$.
    However, an inability to find non-negative integer decompositions in \cref{eq:decompose_symmetry}, is a strict implication of nontrivial band topology, either fragile or stable.
    Because of this, our symmetry-based assessments of topological prevalence in fact represent lower bounds.}.
If the decomposition must include EBRs that are not centered at the unit cell origin, $\mathfrak{n}$ is also said to be an obstructed atomic limit, which may host nonzero bulk polarization or corner charges with corresponding filling anomalies~\cite{vaidya2023topological}.
Conversely, if a decomposition exists with integer $c_{\qv}^{\alpha}$ but requiring at least one negative coefficient, $\mathfrak{n}$ is topologically fragile~\cite{po2018fragile}:
by addition of appropriate atomic limits (namely, the EBRs associated with negative $c_{\qv}^{\alpha}$), the band topology can be trivialized.
In contrast to this, if the decomposition requires non-integer, rational $c_{\qv}^{\alpha}$, the topology of $\mathfrak{n}$ is stably nontrivial~\cite{cano2021band}---being trivializable only by other nontrivial bands.
If no decomposition exists, $\nv$ describes a set of band symmetries that are inconsistent with compatibility relations~\cite{bouckaert1936theory}, \ie a set of bands that are not gapped along high-symmetry $\kv$-lines.
We implemented a software package, MPBUtils.jl~\cite{MPBUtils.jl}, written in the Julia language, to facilitate this analysis, allowing automatic clustering and topological diagnosis of compatibility-respecting photonic bands~%
    \footnote{The 2D PhC problem is spared the complications of the $\Gamma$-point band-symmetry assignment issues of 3D PhCs~\cite{christensen2022location}.
    Instead, the polarization of TE and TM modes imply a natural rule~\cite{de2019engineering}:
    in the zero-frequency at $\Gamma$, the TM (TE) band transform as a $\hat{\mathbf{z}}$-oriented vector (pseudovector).}%
.

\begin{figure*}
    \centerline{%
    \includegraphics[scale=1]{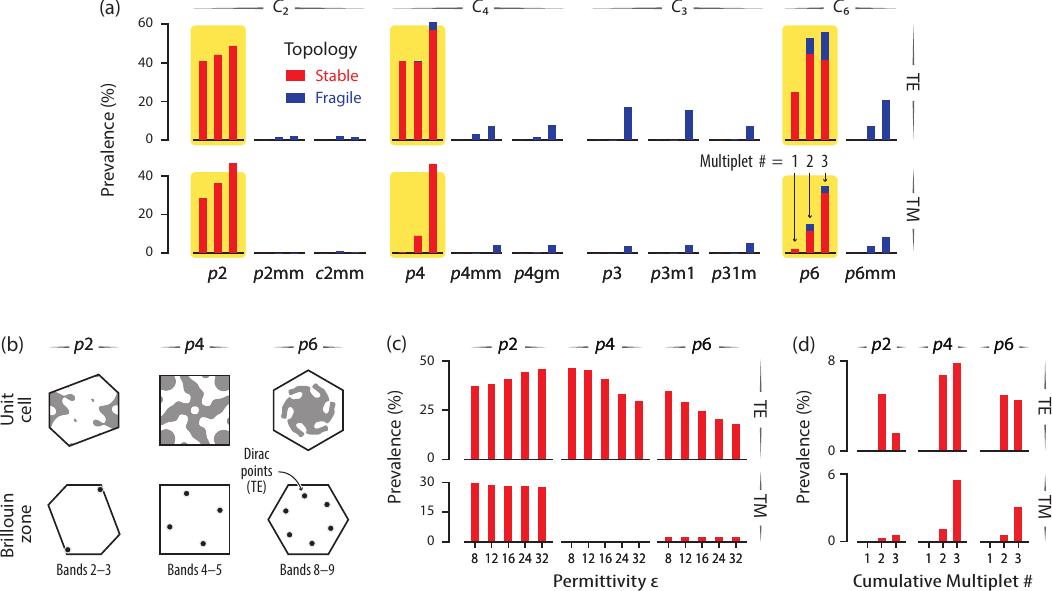}%
    }
    \caption{%
        \textbf{Prevalence of stable and fragile time-reversal invariant topology.}
        (a)~Prevalence of symmetry-identifiable stable and fragile topology across plane groups (permittivity $\varepsilon = 16$; prevalence reported multiplet-by-multiplet).
        Plane groups that allow symmetry-based identification of stable topology are highlighted in yellow.
        (b)~Examples of three PhCs with stable topology and associated Dirac points at generic $\kv$-points (TE polarization).
        The selected examples have no intersecting trivial bulk ``Fermi pockets'', i.e., the Dirac points are frequency-isolated.
        (c)~Permittivity-dependence of stable topology in \emph{p}2, \emph{p}4, and \emph{p}6 for the first multiplet.
        (d)~Fraction of cumulatively stable multiplets whose associated Dirac points are frequency-isolated.
        A near-complete absence is observed in the first multiplet.
        }
        \label{fig:stable-fragile}
\end{figure*}

\paragraph{PhC data set}
Of the 17 plane groups that describe the symmetry settings of 2D PhCs, only 11 allow symmetry-based distinctions between trivial and fragile or stable topology~\cite{po2017symmetry, song2020fragile}, namely those with a proper subgroup of $C_{n\geq2}$.
For each of these 11 plane groups, we created \num{10000} unique, randomly-generated PhC unit cells using a Fourier-based level-set technique~\cite{christensen2022location, christensen2020predictive} (\cref{fig:Workflow}c and Supplemental Section~S10).
Summarizing, we consider a ``two-tone'' dielectric motif in each unit cell: in one region, we place vacuum and in the other a dielectric with permittivity $\varepsilon$, scanning parametrically across $\varepsilon \in \{8, 12, 16, 24, 32\}$, with a filling fraction randomly sampled from a uniform distribution between 0.2 and 0.8.
For each PhC, we compute the requisite band symmetries at high-symmetry $\kv$-points for the first 40 bands, using the MIT Photonic Bands solver~\cite{johnson2001block} and the tooling developed in Ref.~\citenum{christensen2022location}, for both transverse electric (TE) and magnetic (TM) polarizations, obtaining the corresponding symmetry vectors of each separable multiplet (\cref{fig:stable-fragile}ab).
Finally, each multiplet's symmetry-diagnosable band topology is determined via \cref{eq:decompose_symmetry}.
Altogether, our data set encompasses $11$ (symmetry) $\times$ \num{10000} (motif) $\times$ 5 (contrast) $\times$ 2 (polarization) $= \num{1100000}$ distinct calculations.
For a $64\times64$ discretization grid and 40 bands, a typical single-core calculation takes on the order of \qty{20}{\second}.

\paragraph{Stable and fragile topology}
\Cref{fig:stable-fragile} summarizes our results on the prevalence of stable and fragile topology in $T$-invariant 2D PhCs.
We report the prevalence of band topology multiplet-by-multiplet as opposed to cumulatively (\ie up to and including the $n$th multiplet), because cumulative fragile topology is near-absent, as we explain later (Supplemental Section~S6 reports cumulative statistics).

We focus first on stable topology, which is symmetry-diagnosable in just three plane groups---namely, \emph{p}2, \emph{p}4, and \emph{p}6 (\ie with $C_{n\,=\,2,4,6}$-symmetric unit cells)~\cite{po2017symmetry}---and associated with an odd number of Dirac points in each $n$-fold symmetric BZ-sector (equivalently, a $\pi$ Berry phase for loops encircling such sectors)~\cite{fang2012bulk, song2018diagnosis}.
\Cref{fig:stable-fragile}a summarizes our results for a fixed permittivity of $\varepsilon=16$ for the dielectric region.
Surprisingly, stable topology is widely prevalent.
E.g., given a random PhC in \emph{p}2, one should expect Dirac points with a likelihood of $\sim$\,30--50\,\%, depending on multiplet and polarization.
We emphasize that these Dirac points are qualitatively different from the more familiar doublet of Dirac points at the BZ boundary (K and K$'$ points) of $C_6$-symmetric settings~\cite{semenoff1984condensed, ochiai2009photonic}: here, instead, there are $n\bmod 2n$ Dirac points in the BZ interior (\cref{fig:stable-fragile}b).
Polarization-wise, we find that stable topology is more prevalent in the TE than the TM polarization, especially evident in \emph{p}4 and \emph{p}6 where stable topology in the first TM multiplet is near-absent.
Further, varying the permittivity of the dielectric region (\cref{fig:stable-fragile}c), we observe that the overall prevalence is only weakly dependent on dielectric contrast: physically, dielectric contrast mainly controls the size of band gaps---affecting band symmetries and topology only if multiple band closures separate the band structure from the empty-lattice limit.
I.e., geometry is the main design variable for engendering stable PhC topology.
The prevalence of frequency-isolatable Dirac points---\ie Dirac points whose frequency do not intersect other bulk bands---is a small fraction of the overall prevalence of stable topology, however (\cref{fig:stable-fragile}d and Supplemental Section~S12): near-vanishing in the first multiplet and at most a few percent in higher multiplets.

Unlike stable topology, fragile topology is symmetry-diagnosable in all 11 plane groups (\cref{fig:stable-fragile}a).
Although far less prevalent than its stable counterpart, several plane groups feature a substantial prevalence of fragile multiplets, especially among higher multiplets.
We find fragile multiplets in settings with both $C_2T$ symmetry (with relatively winding ``straight'' Wilson loops and a nontrivial Euler class~\cite{cano2021band}) and $C_3$ symmetry (with relatively winding ``concentric'' Wilson loops~\cite{bradlyn2019disconnected, Henke:2022}).
However, we find a near-complete dearth of cumulatively fragile multiplets: across our entire data set, we identify just 43 PhCs with cumulatively fragile topology, all with equivalent EBR decompositions (Supplemental Section S7).
Despite recent progress~\cite{hwang2019fragile, song2020twisted, peri2020experimental, unal2020topological, lange2021subdimensional}, the observable consequences of cumulatively fragile topology remains largely an open question, especially in PhCs~\cite{de2019engineering, alexandradinata2020crystallographic}.
Here, analyzing each cumulatively fragile PhC separately, we find that the corresponding bands feature \emph{pairs} of nodal points in each $n$-fold symmetric sector of the Brillouin zone (Supplemental Section~S7).

\begin{figure}
    \includegraphics[scale=1]{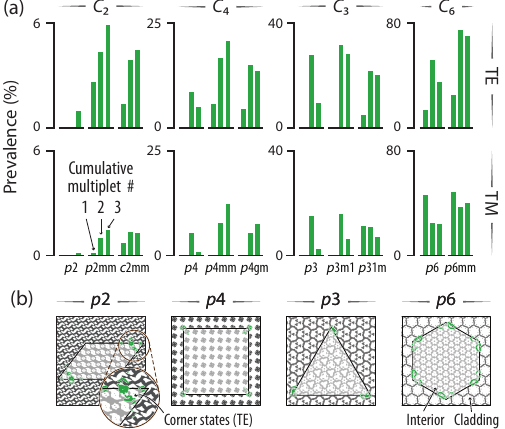}
    \caption{%
        \textbf{Prevalence of corner charges.}
        (a) Prevalence of cumulative higher-order topology, as defined in the main text.
        (b)~Supercell simulations of selected PhC samples with corner charges, cladded by suitable atomic-limit PhCs (Supplemental Section~S13), in \emph{p}2, \emph{p}4, \emph{p}3, and \emph{p}6.
        Contours of energy density shown in green.%
        \label{fig:hoti}
        }
\end{figure}

\paragraph{Higher-order topology}
Beyond the symmetry-indicated classifications of stable, fragile, and trivial (\ie atomic limits) band topology~\cite{kruthoff2017topological, po2017symmetry, bradlyn2017topological} exists a finer gradation associated with higher-order topology and filling anomalies~\cite{benalcazar2019quantization}.
In particular, a filling anomaly exists in any finite $C_n$-symmetric tiling of a PhC unit cell with nonzero bulk polarization ($\mathbf{P}$) or corner charge ($Q$) associated with fractional mode densities at the tiling's edges or corners, respectively.
In turn, $\mathbf{P}$ and $Q$ can be evaluated from band symmetry~\cite{benalcazar2019quantization}---or, equivalently, from the EBR decompositions of \cref{eq:decompose_symmetry} (Supplemental Section S3).

We give values of $\mathbf{P}$ and $Q$ in dimensionless units below, dispensing with the customary-charge prefactor relevant to electronic systems: \ie $Q$ is defined modulo 1 and $\mathbf{P}$ modulo direct lattice vectors.

We restrict our attention to multiplets with nominally trivial cumulative topology, \ie non-negative EBR decomposition coefficients, corresponding to bands that are symmetry-compatible with a gapped atomic limit.
For these multiplets, we classify a given PhC unit cell as higher-order nontrivial if there exists a unit-cell centering choice with vanishing bulk polarization ($\mathbf{P} = \boldsymbol{0}$) and nonzero corner charge ($Q \ne 0$).
This definition is motivated by the following considerations:
(1)~Unlike assignment of stable and fragile band topology, $\mathbf{P}$ and $Q$ are generally dependent on the choice of unit-cell centering (Supplemental Section~S1)~\cite{takahashi2021general}. 
(2)~By requiring $\mathbf{P}=\boldsymbol{0}$ we ensure that a $Q\ne0$ filling anomaly is strictly isolated to corner states, and not coexisting with a corresponding edge anomaly~\cite{schindler2019fractional, vaidya2023topological, benalcazar2017quantized, benalcazar2017electric, benalcazar2019quantization}; additionally, if $\mathbf{P}\ne \boldsymbol{0}$, the definition of $Q$ is dependent on the tiling's boundary configuration in $C_3$-symmetric settings (Supplemental Section~S8).
In practice, for a PhC unit cell with a $C_n$ subgroup symmetry, we evaluate $\mathbf{P}$ and $Q$ for each centering choice that preserves a $C_n$ axis at the unit-cell center, registering the sample's multiplet as higher-order nontrivial if any such centering allows $\mathbf{P}=\boldsymbol{0}$ and $Q\ne0$, jointly with an absence of stable or fragile topology.

\Cref{fig:hoti} reports the prevalence of higher-order topology consistent with this definition, along with supercell calculations of selected samples showcasing the associated corner filling anomaly.
Broadly, prevalence is highest in plane groups with a $C_6$ subgroup, lowest in those with a $C_2$ subgroup, and intermediate those with a $C_4$ or $C_3$ subgroup.
More specifically, we highlight two interesting features in specific plane groups:
First, a complete absence of nontrivial higher-order topology is observed below the third multiplet in \emph{p}2: this is because a $\{\mathbf{P}=\boldsymbol{0}, Q\ne 0\}$ multiplet requires 3 bands in $C_2$- and $C_4$-symmetric settings (Supplemental Section S4) while all \emph{p}2-multiplets are singly degenerate (unlike \emph{p}4, accounting for the nonzero prevalence in its lower multiplets).
Second, in plane groups \emph{p}31m, \emph{p}6, and \emph{p}6mm, there always exists a centering choice that ensures $\mathbf{P}=\bm{0}$ (Supplemental Section S2). Considering this, it is natural to expect a higher prevalence of higher-order topology in these settings.
However, \emph{p}31m mostly defies this expectation, suggesting that the prevalence of higher-order topology in \emph{p}6 and \emph{p}6mm are predominately attributable to their higher rotational symmetry.

\begin{figure}
    \includegraphics[scale=1]{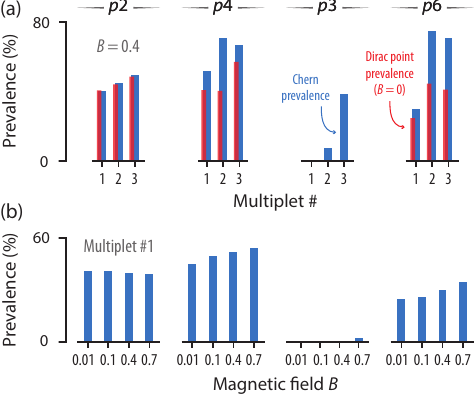}
    \caption{%
        \textbf{Prevalence of Chern topology under time-reversal breaking.}
        (a)~Prevalence of nontrivial Chern topology in the lowest three multiplets an effective magnetic field $B = 0.4$  for TE polarization in plane groups \emph{p}2, \emph{p}3, \emph{p}4 and \emph{p}6. 
        (b)~Prevalence of nontrivial Chern topology in the first multiplet as a function of magnetic field.
        Prevalence is reported multiplet-by-multiplet.
        } \label{fig:chern}
\end{figure}

\paragraph{Time-reversal breaking and Chern topology}
Thus far, we have focused entirely on $T$-invariant PhCs, which necessarily have trivial Chern topology.
To assess the prevalence of nontrivial Chern phases under $T$-breaking, and focusing on the TE polarization~%
    \footnote{The TM modes are unaffected by the considered gyroelectric effect, since their $\mathbf{E}$-fields are $\hat{\mathbf{z}}$-oriented.
    An analogous gyromagnetic effect with nonreciprocal permeability, \ie with $\mu_{xy} = -\mu_{yx}$, breaks $T$ for the TM polarized bands~\cite{wang2008reflection, wang2009observation}.}%
, we incorporate a gyroelectric effect due to a $\hat{\mathbf{z}}$-oriented external magnetic field.
This lifts the scalar permittivity $\varepsilon$ to an anisotropic, nonreciprocal tensor $\boldsymbol{\varepsilon}$ with off-diagonal and diagonal elements $\varepsilon_{xy} = -\varepsilon_{yx} = \iu \varepsilon B$ and $\varepsilon_{xx} = \varepsilon_{yy} = \varepsilon(1 + B^2)^{1/2}$, respectively, where $B$ represents the effective magnetic field amplitude~\cite{lu2013weyl}.

Fixing $\varepsilon=16$ and restricting our attention to the $C_n$-symmetric plane groups---\emph{p}2, \emph{p}4, \emph{p}3, and \emph{p}6---that admit a symmetry-based identification of the Chern number modulo $n$~\cite{fang2012bulk}, we report the prevalence of nonzero Chern numbers in \cref{fig:chern}.
For the first multiplet, the results can be understood (and approximately lower-bounded for small $B$), from the corresponding prevalence of Dirac points under $T$ symmetry (\cref{fig:chern}a, red bars).
Specifically, in 2D, the sources of Chern topology are Dirac~\cite{haldane2008possible} or quadratic degeneracies~\cite{chong2008effective}.
For example, in \emph{p}2, the only such sources are  Dirac points in the BZ interior: as a result, the prevalence of symmetry-identifiable stable $T$-invariant and $T$-broken nontrivial topology agrees exactly at small $B$.
Conversely, \emph{p}4 and \emph{p}6 also allow $T$-protected Dirac and quadratic degeneracies at high-symmetry $\kv$-points, contributing to a higher prevalence of Chern-nontrivial phases than their $T$-invariant counterparts.
Unlike these plane groups, \emph{p}3 does not support robust Dirac points due to its lack of 2-fold rotation symmetry.
Instead, it admits a solitary quadratic degeneracy associated with the $\Gamma_2\Gamma_3$ irrep, whose $T$-breaking can realize nontrivial Chern topology.
The first multiplet of \emph{p}3, however, is singly degenerate and so never includes this irrep, thus explaining the complete absence of nonzero Chern numbers in the first multiplet.

For larger magnetic fields (\cref{fig:chern}b), this simple understanding of breaking $T$-protected degeneracies is modified by the potential for driving multiple band inversions as $B$ is increased.
In the aggregate, however, \ie as an impact on overall prevalence, the influence of $T$-breaking strength appears modest since such inversions tend to produce changes of the band topology that average out over distinct PhC samples.

\paragraph{Discussion}
By comprehensively sampling the space of symmetric 2D PhCs, we have investigated the relative abundance of stable, fragile, and higher-order photonic topology.
Our work elucidates the extent to which various experimental parameters, such as polarization, dielectric contrast, $T$-breaking strength, symmetry, and geometry relate to the prevalence of topological band phenomena.
Overall, we find that this prevalence is affected mainly by polarization, symmetry, and geometry. Contrary to the prevalent association of photonic topology with exotic or rare phenomena---but consistently with recent results from the electronic and phononic domains~\cite{zhang2019catalogue, tang2019comprehensive, vergniory2019complete, vergniory2022all, sodequist2022abundance, xu2022catalogue}---we conclude that photonic topology is very prevalent, even common.
Certain photonic topological features, most notably cumulatively fragile topology, remain rare, however.

Our work motivates future investigations along several lines~\footnote{%
As we were finalizing this manuscript, we became aware of related work under preparation~\cite{huber2023high}.%
}.
For instance, can a partial theoretical understanding be established for the observed statistics, \eg via the empty-lattice approximation?
What is the physical reason for the near-absence of cumulative fragile topology?
And, within a particular symmetry setting, which geometrical motifs are most associated with topological band features?
Our data set~\cite{TopologicalPrevalence} may also serve as the foundations for machine-learning-centric investigations into topological physics~\cite{ma2023topogivity} and photonics~\cite{christensen2020predictive, loh2022surrogate, khatib2021deep}.
Finally, the prevalence observed in our work highlights that the main design challenge of photonic topology is one of finding desired band topology \emph{jointly} with desired spectral properties, \eg large topological gaps or well-isolated degeneracies, motivating further research in topological inverse design~\cite{christiansen2019designing, kim2023automated}.

\vskip 2ex
The data and code used to obtain and analyze the results presented here are made available online~\cite{TopologicalPrevalence}. 

\vskip 2ex
A.G.\ acknowledges the support of the National Science Foundation Graduate Research Fellowship (GRFP).
T.C.\ acknowledges the support of a research grant (project no.~42106) from Villum Fonden.
This research is based upon work supported in part by the Air Force Office of Scientific Research under Grant No.\ FA9550-20-1-0115 and FA9550-21-1-0299, the U.S.\ Office of Naval Research (ONR) Multidisciplinary University Research Initiative (MURI) Grant No.\ N00014-20-1-2325 on Robust Photonic Materials with High-Order Topological Protection, and the U.S.\ Army Research Office through the Institute for Soldier Nanotechnologies at MIT under Collaborative Agreement No.\ W911NF-18-2-0048.
The MIT SuperCloud and Lincoln Laboratory Supercomputing Center provided computing resources that contributed to the results reported in this work.

\bibliographystyle{supp/apsrev4-2-longbib}
\bibliography{main}

\end{document}

%% file: setup.tex
\usepackage[utf8]{inputenc}
\usepackage[english]{babel}

\usepackage{microtype} 
\usepackage{xspace} 

\usepackage{txfonts}  
\usepackage{txfontsb} 

\usepackage{bm} 

\usepackage{xcolor}
\usepackage[]{graphicx} 
\graphicspath{{figs/}}

\usepackage[]{booktabs}
\usepackage{array}
\usepackage{layouts}
\usepackage{multirow}

\usepackage{enumerate}
\usepackage[inline]{enumitem}

\usepackage{xr}
\makeatletter
\newcommand*{\addFileDependency}[1]{
  \typeout{(#1)}
  \@addtofilelist{#1}
  \IfFileExists{#1}{}{\typeout{No file #1.}}
}
\makeatother

\usepackage{hyperref}
\hypersetup{colorlinks,
	linkcolor={blue!75!black!80!yellow},
	citecolor={blue!75!black!80!yellow},
	urlcolor={blue!75!black!80!yellow}
}

\usepackage[capitalize,nameinlink]{cleveref}

\crefname{subequations}{Eqs.}{Eqs.} 
\Crefname{subequations}{Eqs.}{Eqs.}
\crefformat{subequations}{#2Eqs.~(#1)#3}
\Crefformat{subequations}{#2Eqs.~(#1)#3}
\crefname{page}{p.}{p.} 

\usepackage{placeins}

\usepackage{siunitx}
\DeclareSIUnit[number-unit-product = ]\percent{\char`\%} 

\usepackage[centering,hmargin=16mm,tmargin=30mm,bmargin=26mm]{geometry}

\usepackage{soul}

\usepackage{textcomp} 
\usepackage{xifthen}
\usepackage{etoolbox}
\newboolean{togglecomments}
\newboolean{toggletodos}
\newboolean{togglechanges}

\setboolean{togglecomments}{true}
\setboolean{toggletodos}{true}
\setboolean{togglechanges}{false} 

\newcommand{\textblacksquare}{$\blacksquare$}
\newcommand{\swap}[2]{\ifbool{togglechanges}
	{#2}  
	{\textcolor{red!70!black}{[#1]}\textrightarrow{}\textcolor{green!50!black}{[\ignorespaces#2]}}}
\newcommand{\remove}[1]{\ifbool{togglechanges}
	{}    
	{\textcolor{red!70!black}{\ignorespaces#1}}}
\newcommand{\inset}[1]{\ifbool{togglechanges}
	{#1}  
	{\textcolor{green!50!black}{#1}}}
\newcommand{\citeremind}[1]{%
	[\textcolor{blue!75!black!80!yellow}{\textblacksquare%
		\ifthenelse{\isempty{#1}}{}{\textsuperscript{\tiny\textsf{\ignorespaces#1}}}%
	}]\xspace}

\input{commands.tex}

\makeatletter
\newcommand{\raisemath}[1]{\mathpalette{\raisem@th{#1}}}
\newcommand{\raisem@th}[3]{\raisebox{#1}{$#2#3$}}
\makeatother

\renewcommand{\paragraph}[1]{\vskip 1ex\noindent\textbf{#1.}~}

\usepackage[eulergreek]{sansmath}
\makeatletter
\renewcommand\@make@capt@title[2]{%
    \@ifx@empty\float@link{\@firstofone}{\expandafter\href\expandafter{\float@link}}%
    \sansmath\sffamily\textbf{#1\@caption@fignum@sep}#2 
}%

\makeatother


%% file: commands.tex
\newcommand{\qv}{\mathbf{q}}
\newcommand{\kv}{\mathbf{k}}

\newcommand{\nv}{\mathbf{n}}

\newcommand{\iu}{\mathrm{i}}

\newcommand{\ie}{i.e.,\@\xspace} 

\newcommand{\eg}{e.g.,\@\xspace}

\newcommand{\appropto}{\mathrel{\vcenter{
			\offinterlineskip\halign{\hfil$##$\cr
				\propto\cr\noalign{\kern.2pt}\sim\cr\noalign{\kern-2.5pt}}}}}

\DeclareFontFamily{U}{mathx}{\hyphenchar\font45}
\DeclareFontShape{U}{mathx}{m}{n}{<5> <6> <7> <8> <9> <10>
                                  <10.95> <12> <14.4> <17.28> <20.74> <24.88>
                                  mathx10}{}
\DeclareSymbolFont{mathx}{U}{mathx}{m}{n}
\DeclareFontSubstitution{U}{mathx}{m}{n}